\documentclass[twocolumn,nofootinbib,superscriptaddress]{revtex4-1}

\pdfoutput=1

\usepackage{epsf}

\usepackage{epsfig,graphics}
\usepackage {graphicx}
\usepackage {epsfig}
\usepackage {psfig}
\usepackage {subfigure}
\usepackage {tabularx} 
\usepackage{rotate}	
\usepackage{slashed}

\usepackage{amsmath}
\usepackage{amsfonts}
\usepackage{amssymb}
\usepackage{graphicx}

\usepackage{color}
\usepackage[makeroom]{cancel}
\usepackage{graphicx}
\usepackage{amssymb,amsthm}
\usepackage{hyperref}

\usepackage{xcolor}

\def\bi{\begin{itemize}}
\def\ei{\end{itemize}}
\def\be{\begin{equation}}
\def\ee{\end{equation}}
\newcommand{\bea}{\begin{eqnarray}}
\newcommand{\eea}{\end{eqnarray}}

\renewcommand{\Re}{\textrm{Re}\,}

\def\beq{\begin{equation}}
\def\eeq{\end{equation}}
\def\beqa{\begin{eqnarray}}
\def\eeqa{\end{eqnarray}}

\def\Kahler{K\"{a}hler~}

\catcode`\@=11   
\newdimen\@rotdimen
\newbox\@rotbox  

\def\@vspec#1{\special{ps:#1}}%  passes #1 verbatim to the output
\def\@rotstart#1{\@vspec{gsave currentpoint currentpoint translate
   #1 neg exch neg exch translate}}% #1 can be any origin-fixing transformation
\def\@rotfinish{\@vspec{currentpoint grestore moveto}}% gets back in synch 
\def\@rotr#1{\@rotdimen=\ht#1\advance\@rotdimen by\dp#1%
   \hbox to\@rotdimen{\hskip\ht#1\vbox to\wd#1{\@rotstart{90 rotate}%
   \box#1\vss}\hss}\@rotfinish}
\def\@rotl#1{\@rotdimen=\ht#1\advance\@rotdimen by\dp#1%
   \hbox to\@rotdimen{\vbox to\wd#1{\vskip\wd#1\@rotstart{270 rotate}%
   \box#1\vss}\hss}\@rotfinish}%
\def\@rotu#1{\@rotdimen=\ht#1\advance\@rotdimen by\dp#1%
   \hbox to\wd#1{\hskip\wd#1\vbox to\@rotdimen{\vskip\@rotdimen
   \@rotstart{-1 dup scale}\box#1\vss}\hss}\@rotfinish}%
\def\@rotf#1{\hbox to\wd#1{\hskip\wd#1\@rotstart{-1 1 scale}%
   \box#1\hss}\@rotfinish}%
\def\rotate{\@ifnextchar[{\@rotate}{\@rotate[l]}}
\def\@rotate[#1]#2{\setbox\@rotbox=\hbox{#2}\@nameuse{@rot#1}\@rotbox}

\catcode`\@=12
\begin{document}

%----------------------------------------------------------------------%
%  numbering equations with section number
%----------------------------------------------------------------------%
%\makeatletter
%\@addtoreset{equation}{section}
%\makeatother
%\renewcommand{\theequation}{\thesection.\arabic{equation}}
%----------------------------------------------------------------------%
%  title page
%----------------------------------------------------------------------%

\title{Swampland distance conjecture, inflation and $\alpha$-attractors }
\author{Marco Scalisi}
\email{marco.scalisi@kuleuven.be}
\affiliation{Institute for Theoretical Physics, KU Leuven, Celestijnenlaan 200D, B-3001 Leuven, Belgium}
\author{Irene Valenzuela}
\email{i.valenzuela@cornell.edu}
\address{Department of Physics, Cornell University, Ithaca, New York, USA}
\begin{abstract}
The Swampland Distance Conjecture (SDC) constraints the dynamics emerging at infinite distances in field space of any effective field theory consistent with quantum gravity. It provides a relation between the cut-off in energies and the field range which, as we show, in the context of inflation it yields a universal upper bound on the inflaton excursion in terms of the tensor-to-scalar ratio, measured at typical CMB scales.
In this note, we investigate the interplay between the SDC and the emergent inflationary physics around infinite distances singularities in string theory, with a special look at its significance for the $\alpha$-attractor scenario of inflation.  We show that the conjecture itself suggests that inflation may arise as an infinite distance phenomenon with the asymptotic kinetic structure typical of $\alpha$-attractors. Furthermore, we argue that a proper string realisation of these cosmological models in Calabi-Yau manifolds should occur around infinite field distance singularities. However, such constructions typically imply that inflation should not take place in the limit where the inflaton kinetic term develops a pole but rather in the opposite regime. Finally, we study the constraints that the SDC poses on $\alpha$-attractors and show that they still leave considerable room for compatibility with observations.

\vspace{0.2cm}

\end{abstract}
\maketitle

%%----------------------------------------------------------------------%
%%  Paper begins
%%----------------------------------------------------------------------%
%%&&&&&&&&&&&&&&&&&&&&&&&&&&&&&&&&&
\section{Introduction}
%%&&&&&&&&&&&&&&&&&&&&&&&&&&&&&&&&&
In the effort of extracting precise predictions from string theory, it has been noticed that there exist some common patterns, which characterize the string {\it landscape} of consistent effective field theories\footnote{More concretely, these patterns only apply to EFT's weakly coupled to Einstein gravity and that can be UV embedded in a consistent theory of quantum gravity like string theory.} (ETFs).  In contrast, the set of inconsistent EFTs has been termed as belonging to the {\it swampland} \cite{Vafa:2005ui,Ooguri:2006in}. Interestingly, these observations, elevated to conjectures, have triggered the scientific community to investigate their phenomenological implications, which often translate into constraints on the low energy effective theory. In fact, one can show that many seemingly consistent EFTs do not however admit UV completion in string theory. This definitely opens up an exciting avenue towards the possibility of extracting low-energy predictions of quantum gravity.

The Swampland Distance Conjecture (SDC) \cite{Ooguri:2006in} is a proposal for such a quantum gravity constraint. It claims that traversing infinite field distances in string theory always implies the appearance of an infinite tower of particles becoming exponentially light, thus invalidating the EFT. 
This occurs when approaching a boundary of the string moduli space. As long as we stay in this regime, the quantum gravity cut-off of the theory experiences an exponential drop-off in terms of the field distance due to the appearance of the infinite tower of states. This fact automatically translates into an upper bound on the scalar field range $\Delta\varphi$ that an effective theory can accommodate as a function of the quantum gravity cut-off $\Lambda_{\text{QG}}$ in energies, such as 
\beq
\label{bound1}
\Delta\varphi<\frac{1}{\lambda}\log\frac{M_P}{\Lambda_{\text{QG}}}\ .
\eeq
The larger the cut-off, the smaller the field range, so that infinite field ranges become inconsistent with quantum gravity. Strong evidence for the SDC has recently been found in \cite{GPV,Grimm:2018cpv} by going to infinite distance limits in the moduli space of well known string compactifications (see also \cite{Palti:2015xra,Baume:2016psm,Valenzuela:2016yny,Bielleman:2016olv,Blumenhagen:2017cxt,Palti:2017elp,Hebecker:2017lxm,Cicoli:2018tcq,Blumenhagen:2018nts,Buratti:2018xjt,Gonzalo:2018guu}  for previous works and \cite{Lee:2018urn,Lee:2018spm} for a recent analysis in F-theory). The emergent field metric is in fact consistent with the exponential drop-off behaviour of the mass tower predicted by the conjecture. 
Notice that $\lambda$ in eq.~\eqref{bound1} is an unspecified parameter which, in principle, might depend on the type of trajectory followed in the scalar field space. It has been conjectured, though, to be always of order one \cite{Ooguri:2006in,Klaewer:2016kiy} disfavouring very large transplanckian distances (this is known as the Refined Swampland Distance Conjecture \cite{Klaewer:2016kiy}). In \cite{Obied:2018sgi}, the entire rhs of eq.~\eqref{bound1} has been encoded as an order one factor and the corresponding equation denoted as Criterion 1. Among other things, in this note, we aim to clarify the significance of this order one factor as well as the evidence gathered regarding the concrete value of $\lambda$. We will give an overview of this topic in Sec.~\ref{SEC:SDC}.

The SDC can therefore become  a powerful and concrete tool in order to test the regime of validity of ETFs with scalar fields coupled to gravity. The case of cosmological \textit{inflation} with a scalar field crossing a certain distance, in order to deliver around 60 e-foldings of quasi-exponential expansion, is an exemplary situation to investigate. The simple observation that the successful models of inflation should always satisfy $H\leq\Lambda_{\text{QG}}$, with $H$ being the expansion Hubble rate, will allow us to derive a precise upper bound in terms of the tensor-to-scalar ratio, measured at typical Cosmic Microwave Background (CMB) scales. We will discuss this in details in Sec.~\ref{sec:SDCinflation}.

Whereas previous studies have mainly focused on the constraints that the Swampland  imposes when inflation is driven by axionic fields with compact symmetries (see e.g.\cite{ArkaniHamed:2006dz,Rudelius:2014wla,Rudelius:2015xta,Montero:2015ofa,Brown:2015iha,Bachlechner:2015qja,Heidenreich:2015wga} for some pioneering works), in this paper we focus on inflationary models which involve saxions (non-periodic scalars) thus leading to non-compact trajectories (in this case, the uncertainties regarding the value of $\lambda$ are much lower and, in certain cases, it is even possible to give a precise value).  Specifically, we show that the {\it emergent field metric} predicted at infinite distances by the conjecture and confirmed by the asymptotic properties of infinite distance singularities, approaches the form of the one typical of the so-called {\it $\alpha$-attractor} scenario of inflation,
\begin{equation}\label{kin1}
\mathcal{L}_{\text{kin}} = -\frac{3 \alpha}{4\phi^2}(\partial \phi)^2 \,.
\end{equation}
This class of inflationary models has been first proposed in the context of supergravity \cite{Kallosh:2013yoa,Roest:2015qya} but it has been soon realized that its fundamental nature is essentially connected to the form of the inflaton kinetic term \cite{Galante:2014ifa}. The latter induces an attractor for cosmological observables, thus making them insensitive to a wide array of microscopical details which characterizes the theory. The universality regime appears when the scalar potential shows a certain regularity in the limit when the kinetic term shows a pole of order two \cite{Galante:2014ifa,Broy:2015qna,Dias:2018pgj}.

In the present work, we point out that a proper string theory realization of these cosmological models should be engineered when  the inflaton is identified with a scalar field which approaches an infinite distance singularity in field space. This implies, among other things, that the universality regime of $\alpha$-attractor models occurs in the limit where the kinetic Lagrangian eq.~\eqref{kin1} becomes infinitesimally small rather than when approaching the pole. In fact, in Calabi-Yau (CY) compactification manifolds,  the inversion $\phi\rightarrow 1/\phi$ is not necessarily a symmetry and the two scenarios - $\alpha$-attractors and pole-inflation - are not equivalent.  In this context, inflation can be interpreted as an infinite distance emergent phenomenon and the parameter $\alpha$ in eq.~\eqref{kin1}  becomes essentially related to the properties of the singularity and upper bounded by the complex dimension of the CY manifold.  
We will discuss this in Sec.~\ref{Sec:alpha}.

However, approaching infinite distances is not only the limit where we expect the universality of $\alpha$-attractors to emerge but also the limit where the infinite tower of particles becomes exponentially light and the cut-off decreases, signalling the breakdown of the effective theory. It becomes then essential to check consistency of the \mbox{$\alpha$-attractor} models  within the constraints imposed by the SDC. In Sec.~\ref{sec:swampland_const}, we show that  eq.~\eqref{bound1} directly translates into a bound on the total number of e-foldings $N$ during inflation, which is independent of the specific value of $\lambda$ and all the subtleties related to the specific inflaton trajectory. The result is that the EFT of an $\alpha$-attractor model can never support the typical infinite plateau, as expected by consistency with quantum gravity, but the SDC poses nevertheless no restrictions to deliver more than 60 e-foldings of quasi-exponential expansion. Furthermore, we get a relation such as $ \alpha \sim \lambda^{-2}$. Constraints on the value of $\lambda$ will therefore have a direct impact on  $\alpha$-attractor models.

While the above results apply for the case of a single saxion taking large field values, in Sec.~\ref{multifield} we will include some comments about the multi-field case and its implications on the cosmological predictions.

In this note, we keep the focus of our investigation on the kinetic structure of the theory. It is an open question whether the scalar potential can show some  regularity at infinite distances in string theory in order to actually reproduce the cosmological properties of $\alpha$-attractors.  Notice that recent conjectures \cite{Obied:2018sgi,Agrawal:2018own,Ooguri:2018wrx} would disfavour such a scenario. We hope to come back to this issue in the future.

In the following, we will present in detail the arguments in the same order as outlined above and then we will draw our conclusions.

\section{Swampland Distance Conjecture} \label{SEC:SDC}

The Swampland Distance Conjecture, proposed in \cite{Ooguri:2006in}, states that in an effective quantum field theory that can arise from string theory, infinite distances in moduli space lead to an infinite tower of states becoming massless exponentially fast in the proper field distance.  More concretely, if we consider an effective theory valid at a point $Q$ in field space and move to a point $P$, there should appear an infinite tower of states at $P$ with characteristic mass scale $m$ such that 
\be
\label{SC}
\frac{m\left(P\right)}{m\left(Q\right)} \rightarrow e^{-\gamma \Delta\left(P,Q\right)}\quad \mathrm{\;as\;} \quad \Delta\left(P,Q\right) \rightarrow \infty \;,
\ee
where  $\Delta\left(P,Q\right)$ is the  geodesic proper distance between the two points.
Here $\gamma$ is some positive constant which depends on the choice of $P$ and $Q$ but which is not specified in general. A refined version \cite{Klaewer:2016kiy,Agrawal:2018own} of this conjecture implies, however, $\gamma\sim 1$  in Planck units.  The validity of this refined version, even if motivated by plenty of examples in string theory, is under debate and topic of ongoing research. It is also possible to generalise the above conjecture to non-geodesic distances by hiding the path dependence on the value of $\gamma$. We will comment more on this in section \ref{multifield}. 

The key feature of the conjecture is that it predicts the existence of an \textit{infinite} number of particles becoming light. While a finite number of extra new light states would not give necessarily rise to a dramatic change of the theory but to model-dependent corrections, an infinite tower signals the complete breakdown of the effective theory. A quantum field theory description of infinitely many fields weakly coupled to Einstein gravity is no longer possible. One of the consequences of the conjecture is therefore an exponential drop-off of the quantum gravity cut-off as follows,
\beq\label{cutoff}
\Lambda_{\rm QG}=\Lambda_0\ e^{-\lambda \Delta\left(P,Q\right)} 
\eeq
where $\Lambda_0\leq M_P$ is the original naive cut-off of the EFT, and $\lambda\sim \gamma$.

It is also natural to identify this cut-off $\Lambda_{\rm QG}$ with the species scale\footnote{The species scale \cite{ArkaniHamed:2005yv,Distler:2005hi,Dimopoulos:2005ac,Dvali:2007wp,Dvali:2007hz} is given by $\Lambda\sim M_p/\sqrt{\mathcal{N}}$ where $\mathcal{N}$ is the number of light fields or species.} of the tower of particles, as done in \cite{GPV,Heidenreich:2018kpg}, implying $\lambda =\gamma/3$ in four dimensions. The species scale indicates the scale at which quantum gravitational effects cannot be ignored due to the increasing  number of light fields weakly coupled to gravity.  Evidence for this identification can be found in \cite{GPV} and has also been used in \cite{Hebecker:2018vxz}. Notice that this cut-off is conceptually different than the low energy cut-off of the effective theory corresponding to the mass of the lightest particle of the tower. Above the species scale, gravity becomes strongly coupled and any possible quantum field theory description completely breaks down. Furthermore, this drop-off of the quantum gravity cut-off cannot be seen from the point of view of the effective field theory (in the absence of quantum gravity), as expected from any swampland criterium.

\subsection{Evidence at infinite distance singularities}\label{Sec:SDCinfinite}
The SDC finds confirmation in the analysis of the physics around infinite distance singularities (boundaries) of moduli spaces in string theory compactifications \cite{GPV} (see also \cite{Grimm:2018cpv,Lee:2018urn}).  In the following, we will briefly summarise the results and general insights of \cite{GPV}.  First, notice that infinite geodesic distances can only occur when approaching a singularity in the moduli space, where the volume of a cycle goes to infinity or to zero. Interestingly, we can use the theory of limiting Mixed Hodge Structures and the Nilpotent Orbit Theorem by Schmid \cite{schmid} to give the local form of the \Kahler potential (and the field metric) near the singular loci. For Calabi-Yau manifolds, the \Kahler potential near an infinite distance singularity in moduli space located at $\phi\rightarrow \infty$ takes the following form 
\beq
K=-\ln\left(p(\phi)+ \mathcal{O}\left(e^{-T}\right)\right)\,,
\label{Ks}
\eeq
where $T=\phi+i\theta$ is the complex field parametrising the scalar manifold. Here $p(\phi)=\phi^d+\beta\phi^{d-1}+\dots$ is a polynomial depending only on the radial transverse coordinate $\phi$ (the saxion field) approaching the singular point. The degree $d$ of the polynomial is fixed in terms of the properties of the monodromy transformation around the singularity (see \cite{GPV} for more technical details), and it is used to classify the type of singularity. It is always upper bounded by the complex dimension of the Calabi-Yau, $d\leq \text{dim}_{\mathbb{C}}(CY)$ and it is non-vanishing only if the singularity is at infinite distance. Notice that the leading term of the \Kahler potential only depends on the type of singularity and not on the Calabi-Yau in which the singularity is embedded. We can trivially add longitudinal scalars directions $z_a$ along which the singularity expands by allowing the coefficients $\beta_i$ of the polynomial to depend on these fields $z_a$,  i.e. $\beta_i= \beta_i(z_a) $. The generalisation to include more transverse coordinates will be discussed in section \ref{multifield}.

The kinetic term for the field $\phi$, associated to  the \Kahler potential eq.~\eqref{Ks}, reads
\beq
\label{kinSDC}
\mathcal{L}_{\text{kin}} =-\frac{d}{4\phi^2}(\partial \phi)^2
\eeq
which indeed generates a logarithmic divergence of the proper field distance, as it should be when approaching an infinite distance singularity \cite{wang1}.

The tower of states can correspond to a KK tower, winding modes or wrapping branes that become light when moving towards the infinite distance singularity. In any case, using the above parametrization, the mass of the tower of states scales as
\beq\label{massscaling}
m\sim \frac{m_0}{\phi^p}\,,
\eeq
with $p\sim \mathcal{O}(1)$.
%In supersymmetric settings, the effective theory of this field can also be derived from the supergravity low energy effective action given by a Kahler potential of the form $K=-\log(\phi^c+\dots)$. 
Combining eqs.\eqref{kinSDC} and \eqref{massscaling} we can check that the tower of states becomes exponentially light, in terms of the canonically normalised field $\varphi$, such as
\beq\label{massvarphiSDC}
m=m_0\ e^{-\gamma \varphi} \ ,\quad \varphi=\frac{\sqrt{d}}{2}\log\phi
\eeq
as stated by the SDC, with $\gamma=2p/\sqrt{d}$. Notice that the exponential mass behaviour trivially originates from the universal local behaviour of the field metric near infinite distance singularities. More difficult is to check that there are indeed infinitely many particles becoming massless at the singular point. This is clear when thinking of a Kaluza-Klein tower at the large volume point, but becomes highly non-trivial for other infinite distance singularities. In  \cite{GPV,Grimm:2018cpv},  this has been checked for infinite distance singularities at the complex structure moduli space of four-dimensional Type IIB Calabi-Yau compactifications by using the theory of limiting Mixed Hodge Structures at the singular loci. This moduli space presents a very nice and clean realization of these ideas where the massless tower corresponds to BPS states arising from wrapping D3-branes. The same techniques and ideas can also be generalized, though, to \Kahler moduli spaces \cite{pierre}.

Furthermore, there is a natural relation between this conjecture and the absence of global symmetries in this setup. 
Since the leading term of the \Kahler potential \eqref{Ks} does not depend on the axionic partner $\theta$, this field enjoys a global continuous shift symmetry only broken by exponentially suppressed terms. This global continuous symmetry becomes exact at the infinite distance singularity, so the SDC can be understood \cite{GPV} as an obstruction to recover this symmetry by means of an infinite tower of light particles yielding the exponential drop-off of the quantum gravity scale.

Finally, notice that not every singularity is necessarily at infinite distance. If the singularity is at finite distance, the leading polynomial term of the \Kahler potential above vanishes implying that the field metric can never take the form \eqref{kinSDC}. In this case, there is no infinite tower becoming light and, therefore, we do not expect to see an exponential drop-off of the quantum gravity cut-off. Consistently, there is no global continuous symmetry restored at the finite distance singularity.

To summarise, the scenario that seems to emerge underlying the Swampland Distance Conjecture is as follows: The conjecture quantifies how close we can get to the infinite distance point, i.e. how close we can get to the situation of recovering a global symmetry. It does so by providing a relation between the cut-off and the field range due to the appearance of an infinite tower of exponentially massless particles which were not part of the effective theory. The exponential mass behaviour originates from the behaviour of the field metric near the infinite distance singularity, and the \textit{infinite} number of particles implies a drop-off of the quantum gravity cut-off. In other words, above this cut-off quantum gravitational effects cannot be ignored so the model must drastically change (the effects of the tower will never be negligible or subleading). This is very similar to the magnetic Weak Gravity Conjecture (WGC) \cite{ArkaniHamed:2006dz}, which quantifies how small a gauge coupling can be by providing a relation between the gauge coupling and the cut-off scale. This way, the global symmetry limit $g\rightarrow 0$ cannot be reached while keeping a finite cut-off.  Finally, it was also observed in \cite{GPV,Heidenreich:2018kpg} that quantum corrections from integrating out the tower of exponentially massless particles up to the species bound have in fact the structure to generate the infinite field distance. Crucial for this argument is the role of the species bound, which implies an exponentially increasing number of light fields as we approach the infinite distance point. Therefore, the global symmetry and the infinite distance itself might be just emergent phenomena from integrating out infinitely many fields.

\subsection{SDC vs validity of the EFT}

Consider the moduli space of some string compactification. In general, there does not exist a single effective field theory that is  valid globally over the entire moduli space, but we need to work with different effective descriptions which are valid over local patches of finite size. To get these effective theories, we usually expand the field metric and physical observables around special points which correspond to singularities of the moduli space. These singularities can be either at finite or infinite distance. The presence of the latter implies that moduli spaces have finite volume but are non-compact. When we move away from the special point, corrections become more and more important and at some point the local expansion fails. For instance, near an infinite distance singularity (like  large volume, large complex structure point or weak coupling points\footnote{Most of effective theories obtained from string theory compactifications usually involve working near one of these limits. This is why the SDC has the potential to constrain many inflationary models in string theory. There can be, though, other examples of infinite distance singularities.}) the \Kahler potential takes the form given in eq.~\eqref{Ks}. Recall that this is a local expansion around the singularity at $\phi\rightarrow \infty$. Consequently, the exponentially suppressed terms become more important as we move away from the singularity, and at some point (defined as the radius of convergence) the effective description breaks down and needs to be replaced by another one. 

There is some confusion whether this range of validity of the effective theory is the same predicted by the SDC. The answer is no. The SDC is related to a complete breakdown of the effective theory when approaching an infinite distance point (i.e. when moving towards the boundaries of the moduli space), since the quantum gravity \mbox{cut-off} also goes to zero there.  Hence, we cannot just find another quantum effective field theory description (while keeping the same fundamental degrees of freedom)  that works when the first one breaks down, as quantum gravitational effects become important. Furthermore, notice that the SDC implies the breakdown of the effective theory when approaching the special point, instead of when going away from it. In other words, precisely in the limit where the local expansion is better justified, the SDC tells us that the effective theory must break down due to an additional infinite tower of particles becoming light, which was not present in the low energy effective theory. Therefore, as expected from a swampland constraint, this breakdown cannot be seen without additional information of the UV completion. 

The intuitive reason is because there are global symmetries that would be recovered otherwise at the infinite distance boundaries of the moduli space, and global symmetries are not allowed by quantum gravity \cite{Harlow:2018tng}. Therefore, even if near the boundary it seems we can always\footnote{The SDC is deeply linked to the appearance of dualities in string theory.} find a weakly coupled effective theory description enjoying approximate global symmetries and which is seemingly under control from the point of view of QFT, it must still breakdown continuously when approaching the infinite distance boundary by quantum gravity effects to avoid the restoration of the exact global symmetry.
%%%%%%%%%%%%%%%%%%%
\begin{figure}[t]
\begin{center}
\includegraphics[width=6.5cm]{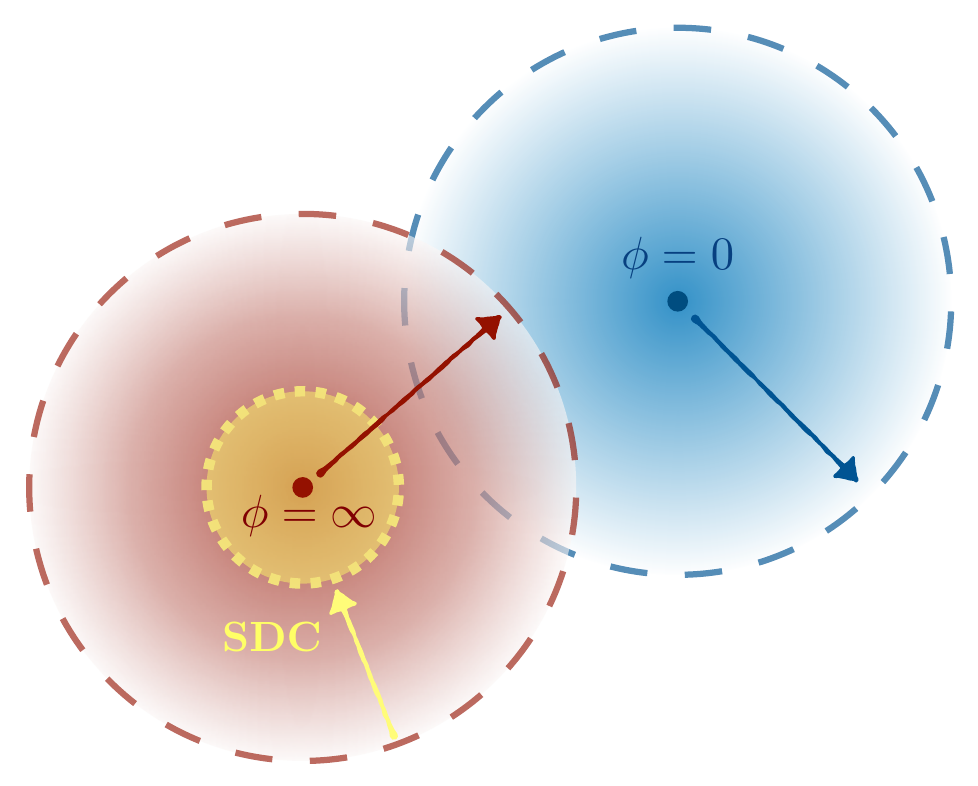}
\end{center}
\vspace{-0.cm}
\caption{\label{gcu}
Sketch of a moduli space with different local patches describing the different regimes of validity of an effective theory. For simplicity, we only draw two singularities, assuming one of them at infinite distance ($\phi=\infty$) and the other one at finite distance ($\phi=0$). The effective field theory is valid at a finite region around the singularity. When going away (red or blue arrow), corrections will become important and eventually has to be replaced by another effective description. Quantifying these corrections defines the radius of validity (dashed circles) of the EFTs. Contrary, the SDC gives information on how close we can go to the singular point (yellow arrow). Even if in this limit the metric at leading order is well described by $1/\phi^2$ and corrections are negligible, there is an additional infinite tower of particles which will yield an exponential drop-off of the energy cut-off. This sets a new boundary (yellow dotted circle) and a corresponding area where the EFT breaks down by quantum gravity effects (yellow area).
}
%\label{plotguay}
\end{figure}

Of course, string theory has a way to resolve these infinite distance singularities (e.g. by growing extra dimensions) so the global symmetry is embedded in a higher group of diffeomorphisms or gauge transformations. But this changes drastically the effective theory and the fundamental light degrees of freedom are intrinsically different. 

Let us finally remark that the effective theory also breaks down at finite distance singularities because of the presence of some new light state. However, since the number of new degrees of freedom is expected to be finite, we can always integrate them in and continue working with the same effective theory plus the new states. There is no need of changing to a dual picture and, consistently, the quantum gravity cut-off does not go to zero in these cases. There have been some works \cite{Blumenhagen:2018nts} studying if a Refined SDC \cite{Klaewer:2016kiy,Obied:2018sgi} forbidding transplanckian geodesic field distances could still be valid in these regimes, even if there is no infinite tower becoming light. They find agreement with the refined conjecture in the sense that the analysed geodesics trajectories are still subplanckian.

\section{SDC and inflation} \label{sec:SDCinflation}

One of the strongest implications of the SDC, and in particular of eq.~\eqref{cutoff}, is a limit on moduli space distances within any effective field theory which is consistent with string theory and has a finite cut-off. Therefore, it is of potential phenomenological interest in the context of inflation where both a high energy scale and a large field excursion may play an important role. 

In order to have a successful inflationary model, we need the cut-off to stay above the Hubble scale, that is 
\begin{equation}
H\leq \Lambda_{\rm QG}\,.
\end{equation}
This simple observation allows us to give a {\it model-independent upper bound} on the proper field distance $\Delta\varphi$ that any inflationary model (consistent with quantum gravity) can accommodate
\beq
\label{bound}
\Delta\varphi<\Delta\varphi_{\text{SDC}} =\frac{1}{\lambda}\log\frac{M_p}{H}\,,
\eeq
assuming that the conjecture is valid for any scalar field taking parametrically large values\footnote{More concretely, it applies to EFTs where the limit $\Delta\varphi\rightarrow \infty$ corresponds to approaching an infinite distance boundary of the moduli space, as in most of the string theory realizations of large field inflationary models.}.
Recalling that in the slow-roll limit we have
\beq
\frac{M_p}{H}=\sqrt{\frac{2}{\pi^2 A_s\ r}}\,,
\eeq
the bound eq.~\eqref{bound} can be expressed  in terms of the tensor-to-scalar ratio $r$
\beq
\label{boundhorizon}
\Delta\varphi<\Delta\varphi_{\text{SDC}} =-\frac{1}{2\lambda}\left(\log\frac{\pi^2 A_s}{2}+\log{r}\right)\,,
\eeq
with $A_s$ being the amplitude of scalar perturbations. Notice that the quantities $H$, $A_s$ and $r$ are calculated at the scales which crossed the Hubble horizon around 60 e-foldings before the end of inflation (the same scales are entering our present horizon and correspond to the largest ones of our observable Universe). These are the only scales of interest for the current analysis as they are associated with the highest inflationary energy we can effectively probe through cosmic microwave background (CMB) experiments, such as the Planck satellite \cite{Akrami:2018odb,Aghanim:2018eyx}. For the sake of simplicity, here and in the following, we do not label these quantities with a star, as it is usually done in literature.

\begin{figure}[t]
%\vspace*{3mm}
%\hspace{-3mm}
\begin{center}
\includegraphics[width=8.5cm]{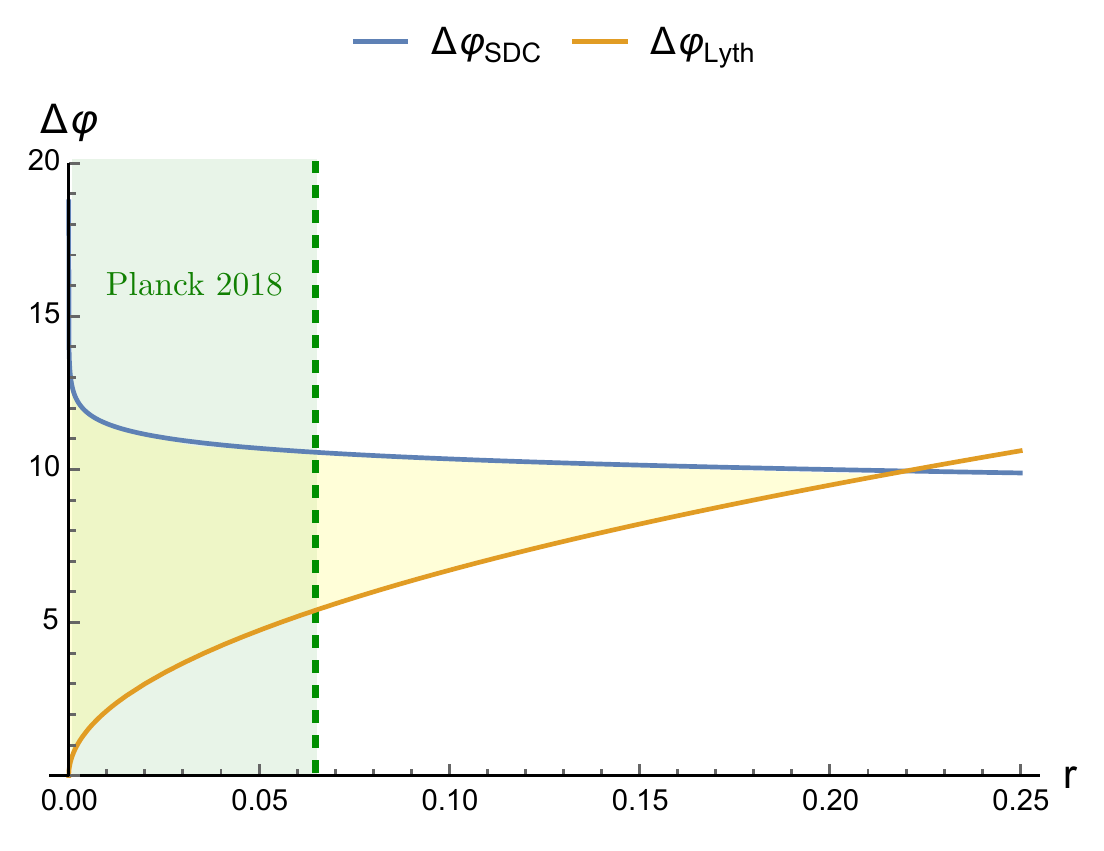}
\end{center}
%\vspace*{-0.3cm}
\caption{The blue line is the universal upper bound predicted by the SDC as function of the tensor-to-scalar ratio $r$ (we have assumed $\lambda=1$). The orange line is the Lyth bound which provides a minimum range for generic models of inflation. Scenarios of inflation whose EFT is consistent with quantum gravity should have a field range which belongs to the yellow area. The green area is the one allowed by Planck 18, via the upper bound $r\leq 0.064$ (dashed green line).}
\label{FIG:bounds}
\end{figure}

First of all, let us  note that the lower the inflationary scale is, the milder the bound becomes. This means that the EFT of models with a small Hubble parameter, which generically correspond to scenarios where a sub-Planckian field excursion is sufficient to generate 50-60 e-foldings of inflation, will be safe in this respect. The point in field space where the EFT is supposed to break down reduces to smaller values together with the increase of the inflationary energy scale, making large-field models more in tension with such a conjecture. This behaviour is shown in fig.~\ref{FIG:bounds} by the blue line, which represents the bound eq.~\eqref{bound} for $\lambda=1$.

The latest Planck results \cite{Akrami:2018odb}, through the upper bound on the tensor-to-scalar ratio (dashed green line in fig.~\ref{FIG:bounds}), impose
\beq \label{Hbound}
\frac{M_p}{H}>3.7\times10^4\,,
\eeq
which, once plugged into eq.~\eqref{bound}, translates into a maximum field excursion set by the SDC of around 10 $M_p$ (when $\lambda=1$). Interestingly, given the measured value of the scalar amplitude $A_s$ \cite{Aghanim:2018eyx}, the main contribution is given by the first addend in eq.~\eqref{boundhorizon} which alone provides a super-Planckian excursion of around  9.19 $M_p$ ($\lambda=1$). A smaller value of $\lambda$ would allow to traverse a larger field displacement before the breakdown of the EFT of inflation\footnote{There might be some room to make $\lambda$ smaller by travelling along non-geodesic trajectories in a multi-field scenario. This is discussed in section \ref{multifield}.}.

The bound is therefore yielding a maximum field excursion which is roughly $\mathcal{O}(1)$ in reduced Planck mass units (we note that while this is a consequence of the Refined SDC \cite{Klaewer:2016kiy}, it was further emphasized as a swampland criterion in \cite{Agrawal:2018own}), but it can still leave considerable room for inflationary models with $\Delta\varphi>1$ allowed by the current data. In fact, given the latest cosmological constraints provided by Planck \cite{Akrami:2018odb}, we expect at most modest super-Planckian excursions and, therefore, determining the actual value of $\lambda$ becomes of crucial importance in order to determine any possible tension with the limit imposed by the SDC. Furthermore, we would like to emphasize that this conjecture does not provide a universal strict bound on $\Delta\varphi$ but rather a relation between $\Delta \varphi$ and $\Lambda_{\rm QG}$, which are two quantities in principle disconnected in the absence of gravity. This can have important implications for inflation, besides the constraint on the field range, due to the premature breaking down of the EFT.

It is interesting to compare the upper bound eq.~\eqref{bound} with the famous Lyth bound \cite{Lyth:1996im,Boubekeur:2005zm,Garcia-Bellido:2014wfa}, which provides a minimum for the field excursion of slow-roll inflationary models\footnote{The original bound derived by Lyth  \cite{Lyth:1996im} was taking into account just the small observable window which CMB experiments have access to. Assuming that $r$ monotonically increases as inflation proceeds, one can extend this bound to the whole period of inflation ($N\simeq 60$).}. Whereas the Lyth bound increases together with $r$, the SDC bound decreases and the two functions define a finite area of validity for inflationary models consistent with quantum gravity (see yellow area in fig.~\ref{FIG:bounds}). It is interesting to notice that many models of inflation, which have been ruled out by the cosmological data of the last couple of years, are also in very strong tension with quantum gravity arguments. A primary example of this situation is provided by the simple quadratic model \cite{Linde:1983gd} of inflation (with $r\sim0.1$ and $\Delta\varphi\sim 15$), which is excluded by the swampland distance conjecture for $\lambda\geq1$.

Let us finally remark that the infinite tower of states is not part of the effective field theory and can only be identified if one has a UV completion in a consistent theory of quantum gravity. Therefore, the above bound could have never be obtained by studying the validity of the model within the effective field theory itself, without extra input about the tower of states. Furthermore, the bound disappears when decoupling gravity by sending $M_p\rightarrow \infty$, as any swampland constraint should.

\section{$\alpha$-attractors and string theory}\label{Sec:alpha}
\subsection{Review: $\alpha$-attractors = pole inflation}
The $\alpha$-attractors cosmological scenario  \cite{Kallosh:2013yoa,Roest:2015qya} has been proposed and developed in the framework of supergravity (see also \cite{Kallosh:2013hoa,Ellis:2013nxa,Roest:2013aoa} for previous investigations with some working examples for specific values of $\alpha$) and subsequently studied in a variety of contexts. The crucial observation is that the kinetic structure of the theory, and therefore the underlying \Kahler geometry, may act as an ``attractor'' thus determining the inflationary observables unequivocally, irrespective of a certain array of details. Specifically, the predicted values for the scalar spectral index $n_s$ and the tensor-to-scalar ratio $r$, measured by CMB experiments, have universal form
\begin{equation}\label{nsr}
n_s= 1-\frac{2}{N}\,, \qquad r=\frac{12\alpha}{N^2}\,,
\end{equation}
at large values of the number of e-foldings $N$ and  with $\alpha$ being a numerical parameter directly related to the curvature of the scalar manifold. 

The key ingredient of this universality is a dependence of the \Kahler metric on the inflaton $\phi$ such that 
\begin{equation}\label{kin}
\mathcal{L}_{\text{kin}} = -\frac{3 \alpha}{4\phi^2}(\partial \phi)^2 \,.
\end{equation}
This situation appears for typical logarithmic \Kahler potentials such as $K=-3\alpha\ln(T+\bar{T})$ when the inflaton is identified with the saxion, that is $\phi \equiv \Re T$.

This class of models is therefore characterized by an underlying {\it scale invariance} \cite{Csaki:2014bua,Ozkan:2015kma} which can be broken by means of various mechanisms ({\it e.g.} by a field-dependend term in the superpotential \citep{Kallosh:2013yoa,Roest:2015qya,Scalisi:2015qga}, by K\"ahler \cite{McDonough:2016der}, loop~\cite{vonGersdorff:2005bf,Berg:2005ja,Berg:2007wt,Cicoli:2007xp,Cicoli:2008gp} or higher-derivative~\cite{Ciupke:2015msa,Broy:2015zba} corrections) thus generating an inflationary dynamics. In full generality, the universal cosmological predictions arise when the scalar potential is regular at {\it small-distance} in field space and can therefore be expanded as
\begin{equation}
\label{Vsmall}
V= V_0 - V_1\ \phi + \mathcal{O}(\phi^2) \,,
\end{equation}
where  the kinetic term eq.\eqref{kin} shows a pole \cite{Galante:2014ifa} (indeed, this class of models is often termed {\it pole inflation} \cite{Broy:2015qna,Dias:2018pgj}). This regularity translates into an infinite positive plateau with a first exponential deviation in terms of the canonically normalized variable $\varphi=-\sqrt{3\alpha/2}\ln\phi$. However, in order to solve the standard cosmological puzzles, one needs just a finite amount of inflation (usually quantified with 50-60 e-foldings). Therefore, the observational predictions would not be affected if this regularity holds just for a certain finite field range, while allowing even singular behaviour in the limit $\phi\rightarrow 0$. An interesting example of this situation is the fiber inflation scenario \cite{Cicoli:2008gp}, where power corrections to the originally flat direction enters the potential with both signs, thus spoiling the infinite quasi-de Sitter phase at very large field values.

It is therefore useful to give an expression of the canonical field range in terms of the number of e-foldings $N$. This has been derived in \cite{Garcia-Bellido:2014wfa} and reads
\begin{equation}
\label{frange}
\Delta\varphi=\sqrt{3\alpha/2}\ln N-\varphi_e \,,
\end{equation}
where $\varphi_e$ is the point where inflation ends and it is usually a sub-leading contribution at large values of $N$. As an example, models with $\alpha=1$ (including the Starobinsky model \cite{Starobinsky:1980te}) will deliver 60 e-foldings of inflation when the scalar field moves of about 5 $M_p$. Therefore, modifications or breakdown of the EFT beyond this value will not affect the original inflationary predictions. The above expression can be obtained by integrating the relation $d \varphi/ dN = \sqrt{2 \epsilon}$, with the first slow-roll parameter being equal to $\epsilon=3\alpha/4N^2$, at leading order in $N$.

In the following, we would like to argue that a proper realization of $\alpha$-attractors in string theory corresponds to taking the opposite limit $\phi \rightarrow \infty$, that is the limit where the kinetic term goes to zero and then has no pole.

\vspace{0.5cm}

\subsection{String realisation of $\alpha$-attractors\\ at infinite distance ({\it i.e.} $\alpha$-attractors $\neq$ pole inflation) }

Let us start with the observation that the key-ingredient of $\alpha$-attractors, namely the kinetic term eq.~\eqref{kin}, has the same $\phi$-dependence of the metric eq.~\eqref{kinSDC} appearing only when approaching an infinite distance singularity in field space. This suggests a very natural realization of $\alpha$-attractors at the boundaries of the string moduli space. Here below, we discuss this possibility and highlight the differences and possible issues with the original pole-inflation scenario described above. 

Let us first note that the kinetic term eq.~\eqref{kin} is invariant under the inversion  $\phi\rightarrow1/\phi$. This implies that the phenomenology of $\alpha$-attractors remains the same if the scalar potential can be expanded at {\it infinite distance} such as\begin{equation}
\label{Vlarge}
V= V_0 - V_1 /\phi + \mathcal{O}(1/\phi^2) \,.
\end{equation}
Note that, after the inversion, the kinetic term  eq.~\eqref{kin} will have no longer a pole in the deep inflationary phase limit, rather will be infinitesimally small. Again, also in this parallel case, positive power corrections might enter the above expression, while still yielding \mbox{$\alpha$-attractor} behaviour at infinite distance, if the plateau extends at least for 50-60 e-foldings.

While this discussion is correct at the level of the effective theory of inflation with a single real scalar field, one has to be careful when embedding these models in string theory, which necessarily implies new degrees of freedom coupled to $\phi$. 

Let us consider a simple supergravity embedding, in which the inflaton $\phi$ is identified with the real part (saxion) of one complex field $T$  parametrising a scalar manifold with \Kahler potential 
\beq\label{Kalpha}
K=-3 \alpha\ln(T+\bar{T}+\ldots)\,.
\eeq
This type of \Kahler potential arises when working in a local region near an infinite distance singularity (like e.g. \textit{large volume} or \textit{large complex structure}). As mentioned earlier in Sec.~\ref{Sec:SDCinfinite}, we usually have only a local patch-wise description of the moduli space of a string compactification.
The latter expression should indeed be understood as an expansion around an infinite distance singularity  where we are neglecting sub-leading corrections denoted by the ellipsis.  The local form of the \Kahler potential near infinite distances singularities was given in eq.~\eqref{Ks}. Comparing both equations, it is trivial to check that eq.~\eqref{Kalpha} is a particular case of eq.~\eqref{Ks} with $p(\phi)=(T+\bar{T})^{3\alpha}=(2\phi)^{3\alpha}$.

If we now perform an inversion transformation such as $T\rightarrow T' =1/T$, the leading term of the \Kahler potential eq.~\eqref{Kalpha} remains invariant (up to a \Kahler transformation). This implies that the kinetic Lagrangian
\beq\label{K}
\mathcal{L}_{\text{kin}} =-3 \alpha \frac{\partial T \partial \bar{T}}{(T+\bar{T})^2}= -\frac{3 \alpha}{4\phi^2}\left((\partial \phi)^2+(\partial \theta)^2\right)\
\eeq
is covariant with respect to the inversion, so that  \mbox{$\mathcal{L}_{\text{kin}}(T)= \mathcal{L}_{\text{kin}}(T'=1/T)$}. In terms of the new field \mbox{$T'=\phi'+i\theta'$}, the kinetic term for $\phi'$ looks the same but the potential changes. Assuming that $\phi$ enjoys a potential of the form eq.~\eqref{Vlarge}, the potential in terms of $\phi'$ will look as in eq.~\eqref{Vsmall}. This implies that the inflationary regime emerges for $\phi\rightarrow \infty$, whereas in the primed frame it does for $\phi'\rightarrow 0$. At this point, it is interesting to notice that, since the field metric is covariant under the inversion transformation, the decay constant of $\theta$ goes to zero as $\phi\rightarrow \infty$, while the decay constant of $\theta'$ diverges as $\phi'\rightarrow 0$. This happens in the same physical situation (\mbox{{\it i.e.}} during inflation). However, only $\theta$ can be identified with the {\it true} axion, as we explain in the following. 

In the new primed frame, the separation on $K$ between a leading polynomial (depending only on the radial variable) and the exponentially suppressed corrections is no longer true (this is generically the case for prototypical four-dimensional Calabi-Yau compactifications of Type II string theory). In particular, the angular variable $\theta'$ is not an axion in the sense that it does not enjoy an approximate continuous shift symmetry only broken by exponentially suppressed corrections.  Therefore, even if its field metric diverges, this does not have the physical meaning of a ``decay constant" and  it does not make sense to apply the bounds derived by the Weak Gravity Conjecture. The true axion is only $\theta$, whose decay constant goes to zero at the infinite distance singularity. Therefore, a kinetic term of the form \eqref{K}, where it is assumed that the corrections are exponentially suppressed so that $\theta$ behaves as an axion, is only valid for  $\phi\rightarrow \infty$. 

The inversion transformation should not be understood as a symmetry but more as a duality (two different descriptions giving the same physics). It is part of the group of linear transformations $GL(2n+2,R)$ which leaves invariant the \Kahler potential  up to \Kahler transformations. However, in general, these transformations do not correspond to symmetries of the effective theory but rather to a choice of frame (field redefinitions or duality transformations). The true symmetries of the action are a discrete subgroup $M\in GL(2n+2,R)$ called the monodromy group, which does not include in general this inversion transformation. This is more clear if the scalar fields are part of $N=2$  vector multiplets together with $U(1)$ gauge fields (as in the complex structure (K\"ahler) moduli space of IIB (IIA) Calabi-Yau compactifications). The scalar field metric is then related to the gauge coupling, and the inversion transformation implies in turn a transformation on the gauge coupling $g\rightarrow 1/g$ and the exchange between electric and magnetic gauge fields. At a given point of the moduli space, an effective description in terms of electric gauge fields is dual to a description of the magnetic gauge variables, but if the electric gauge coupling goes to zero, the magnetic one will diverge and there is no weakly coupled effective Lagrangian that we can write for the magnetic variables. It is more useful then to work with the electric fields. Similarly, for the scalar manifold, there are frames which are more useful than others depending on the point of the moduli space. When approaching an infinite distance singularity, there is a ``clever'' frame for the \Kahler potential in which the corrections are exponentially suppresed with respect to the leading term, so the shift symmetry of the axion is manifest. But in this frame, the singularity is located at $\phi\rightarrow \infty$ so the result is only valid for large $\phi$. The Nilpotent Orbit Theorem naturally selects this ``clever'' frame and gives the leading form of the \Kahler potential at any type of infinite distance singularity, regardless of the specific Calabi-Yau compactification space or the scalar field under consideration.

Notice that there can be exceptions for particular scalar manifolds in which the inversion transformation is indeed a true symmetry of the theory, as in toroidal compactifications. There, the monodromy group is $SL(2,Z)$ and both limits $\phi\rightarrow \infty$ and $\phi\rightarrow 0$ correspond to infinite distance singularities.  For Calabi-Yau compactifications, only special examples exhibit infinite distance singularties at small $\phi$ which are in fact characterized by small values of $d$ in eq.\eqref{kinSDC} (in particular, $d<3$ for Calabi-Yau threefolds \cite{GPV}) which imply in turn small values of $\alpha$.  Clearly, one can always choose to work in the frame in which the singularity is located at infinity, so the presence of the axion is manifest and everything works as described above. However, in these cases, both $\theta$ and $\theta'$ behave as axions in the effective theory. 

To sum up, the \Kahler potential in eq.~\eqref{Kalpha} is a local expansion at $\phi\rightarrow \infty$ and, in general, it is inconsistent to work with this effective theory when moving to the pole $\phi\rightarrow 0$ and assume that the corrections will be exponentially suppressed and remain negligible. Furthermore, in \Kahler manifolds, each infinite distance singularity implies the presence of an axion in the effective theory with an approximate continuous shift symmetry, and whose decay constant goes to zero at the singularity. This behaviour for the decay constant  holds for any infinite distance point that belongs to a single singular divisor in the moduli space, meaning that we are sending only one field $\phi$ to infinity. As we will comment in section  \ref{multifield}, the case of sending several fields to infinity is more technically involved and has not been proven yet, although we still expect it to be true. We hope to come back to a more detailed analysis of the field metrics in this case in the future. Notice that fiber inflation \cite{Cicoli:2008gp} enters in this category, as two \Kahler fields are sent to large values simultaneously. 

Finally, as in any inflationary model, it is important to discuss the robustness of the effective theory against higher order corrections. The separation in the \Kahler potential between the leading polynomial and the exponentially suppressed corrections allows for the  flattening of the scalar potential if $V$ takes the form \eqref{Vlarge}. Therefore, as far as we can see, only if the inflaton is identified with a scalar field approaching an infinite distance singularity,  the characteristic \emph{plateau} of $\alpha$-attractors can be safely generated. However, this might not be sufficient. The advantage of identifying the inflaton with an axion is the presence of an approximate continuous global symmetry which protects the scalar potential from higher order corrections. This shift symmetry is inherited from the monodromy transformation of infinite order present at any infinite distance singularity, which implies that the axion does not appear in the \Kahler potential up to exponentially suppressed corrections. It is then natural to ask whether there is any analogous protection for a saxion field. Even if this is generically not possible, the analogous protection for the saxion would come from an approximate scaling symmetry, i.e. $\phi\rightarrow \xi\phi$ with $\xi$ being a constant, preserved at leading order in the \Kahler potential.  This scaling symmetry indeed corresponds to a shift symmetry for the canonically normalised saxion \cite{Csaki:2014bua,Carrasco:2015uma}. Unlike the axionic case, this scaling symmetry will be related to the finite order part of the monodromy transformation. Whether the monodromy transformation associated to an infinite distance singularity can indeed generate this scaling symmetry, is not clear and deserves further investigation.

A last cautionary remark regarding the scalar potential is in order. In this note, we only focus on the kinetic structure of the inflaton. Therefore, the proximity to these types of singularities is a necessary but not sufficient condition to get the cosmological properties of \mbox{$\alpha$-attractors}. %It seems hard to get a constant term $V_0\neq 0$ at an infinite distance singularity, as typical examples point either to a divergence or a vanishing of the scalar potential at infinite distance. However, we cannot discard this possibility yet if approaching infinite distance multi-parameter singularities with small values of $d$ (so they do not correspond to the large volume or large complex structure points). %Notice, though, that an exponential runaway would be in agreement with recent conjectures \cite{Obied:2018sgi,Agrawal:2018own,Ooguri:2018wrx,Hebecker:2018vxz}.
 The analysis of the asymptotic behaviour of the scalar potential is left for future work.

\section{Swampland constraints on $\alpha$-attractors\label{sec:swampland_const}}

The Swampland Distance Conjecture suggests that the effective theory of any scalar field\footnote{We will comment on the generalisation to axions in section \ref{multifield}.}, when going to parametrically large enough values, exhibits a universal behaviour for the kinetic metric. This is also confirmed by the analysis of the physics around infinite distance singularities \cite{GPV}. % for the radial paths approaching the singularity, and gives an appealing stringy motivation for these models. 
Interestingly, as we have pointed out in the previous section, this emergent field metric eq.~\eqref{kinSDC} precisely matches the one required for $\alpha$-attractors eq.~\eqref{kin}. By comparing both equations, we are then encouraged to identify
\begin{equation}\label{dalpha}
d=3\alpha\,.
\end{equation}
Deviations from the perfect $1/\phi^2$ dependence, as it is in eq.~\eqref{kin}, may certainly occur but this would affect the inflationary dynamics just at smaller values of the field and, then, far from the CMB window where the observational predictions conform to eq.~\eqref{nsr}.  If the scalar potential takes the form \eqref{Vlarge} with $V_0\neq 0$, the dynamics will eventually be characterized by a long enough plateau and an inflationary $\alpha$-attractor behaviour. Contrary, if $V_0=0$, there will be an exponential runaway towards large field values, which might be useful for quintessence. This latter behaviour would be in agreement with recent conjectures \cite{Obied:2018sgi,Agrawal:2018own,Garg:2018reu,Ooguri:2018wrx,Hebecker:2018vxz}.

The universality features of $\alpha$-attractors occur at large values of $\phi$, i.e. where the scalar potential develops a quasi-de Sitter plateau in canonical coordinate $\varphi$. However, this is also in the same limit where the infinite tower of states predicted by the SDC becomes light and the effective theory breaks down. It is therefore essential to check how far we can move along the plateau (towards the infinite distance point) before the quantum gravity cut-off decreases such that becomes lower than the Hubble inflationary scale.

\subsection{Universal bound independent of $\lambda$ }
The maximum field range that can be traversed before the effective theory breaks down is given by eq.~\eqref{bound}. We can combine the latter with eq.~\eqref{frange} in order to obtain
\beq
\Delta\varphi=\sqrt{3\alpha/2}\log N\leq \frac{1}{\lambda}\log\frac{M_p}{H}\,,
\eeq
which translates into a bound on the total number of e-foldings
\beq
N\leq \left(\frac{M_p}{H}\right)^{\sqrt{\frac{2}{3\alpha}}\frac{1}{\lambda}}\,.
\eeq

Let us recall that the parameter $\lambda$ is a combination of the coefficient $d$ appearing in the field metric eq.~\eqref{kinSDC} and the power of the mass $p$ for the tower of states eq.~\eqref{massscaling} so that $\lambda=2p/(3\sqrt{d})$. Upon using eq.~\eqref{dalpha}, we then obtain\footnote{A slightly stronger bound can be obtained if we impose that the lightest state of the infinite tower remains always heavier than $H$, such that the flatness of the inflationary potential is preserved. In this case we get
$
 N\leq \left(m_0/H\right)^{\frac{1}{\sqrt{2}}\frac{1}{p}}
$
where $m_0$ is the mass of the lightest state at the furthest point from the singularity during inflation, that is where the inflationary dynamics ends. This bound is however more model-dependent, thus we prefer to use the universal criterion given by the species bound.}
\beq \label{Nbound}
 N\leq \left(\frac{M_p}{H}\right)^{\frac{3}{\sqrt{2}}\frac{1}{p}}\,.
\eeq
This bound is universal for any $\alpha$-attractor model and with a very mild dependence on $\alpha$ through the energy scale ($H\propto \sqrt\alpha$). There is still some dependence on $p$, which is an order one factor appearing in eq.~\eqref{massscaling} and which depends on the specific structure of the tower of particles. Typically, $p$ is a (half)-integer upper bounded by the complex dimension of the internal space, i.e. $p=1/2,1,\dots,3$ for a $CY_3$, depending whether the tower of particles arises from wrapping branes of different dimensionality, KK modes, winding modes, etc. 

Clearly, the inequality eq.~\eqref{Nbound} easily holds for typical inflationary periods of 60 e-foldings, given the current experimental bound on the Hubble scale eq.~\eqref{Hbound} coming from the constraints on tensor modes.  In the next subsection, we will get a stronger bound by fixing the value of $\lambda$ in concrete scenarios.  It is worth to remark, though, that eq.~\eqref{Nbound} is a universal upper bound for $\alpha$-attractor models obtained by requiring consistency with quantum gravity via the SDC, regardless of all the subtleties related to the exact value of $\lambda$ and independent of whether the field range is transplanckian  or not. Furthermore, since the dependence on $d$ cancels out, this bound equally applies to the higher-dimensional moduli spaces and any type of trajectory approaching an infinite distance point.

\subsection{Bounds on $\lambda$, $\alpha$ and  $r$ \label{bounds}}
 We can now be more restrictive and comment on the allowed range for $\lambda$. This is not specified by the conjecture, although  a refined version of the SDC states that $\lambda$ should always be of order one for geodesic trajectories. This would automatically imply that $\alpha< \frac{2}{3}\mathcal{O}(1)$ and $\Delta\varphi<\log(\frac{M_p}{H})$. Interestingly, we can give more quantitative bounds if using the results for the geometry near infinite distance singularities of the moduli space. In particular, in certain cases, we can fix  this order one factor in terms of the type of singularity arising at $\phi\rightarrow \infty$. 

In this section, for simplicity, we will only consider single field inflationary models in one dimensional moduli spaces. We will comment on the generalization to more realistic higher dimensional moduli spaces and more general trajectories in section \ref{multifield}. For single field, the factor $d$ appearing in the field metric \eqref{kinSDC} is simply an integer that characterises the type of infinite distance singularity and  is given by the properties of the monodromy transformation around the singularity\footnote{More concretely, $d$ is an integer corresponding to the maximum power of the nilpotent monodromy operator that does not annihilate the period vector. More details can be found in \cite{GPV}. }. It is upper bounded by the complex dimension of the compactification space (for a Calabi-Yau threefold, $d=1,2, 3$). Then, eq.\eqref{dalpha} implies $\alpha=1/3,2/3,1$ respectively. The maximum value $d=3$ ($\alpha=1$) corresponds to a maximal unipotency singularitiy, like the large complex structure point (or large volume). This implies that, $\alpha$-attractor models described by a single scalar field approaching an infinite distance singularity in a one dimensional moduli space of Calabi-Yau compactifications of Type II string theory have necessarily $\alpha \leq 1$, implying
\beq
\label{boundr}
r= 12\alpha/N^2\leq 0.003\,,
\eeq
for 60 e-folds of inflation. This bound is valid for any Calabi-Yau manifold, but restricted to a purely saxionic trajectory approaching the singularity. We will comment on more general trajectories in the next section.

\section{Comments on the multi-field case\label{multifield}}

The generalisation of the SDC to more than one scalar field is subject of controversy and open questions. Similar to the case of a single scalar, one would expect that, as long as the trajectory in the scalar manifold approaches an infinite distance point, an infinite tower of states will become exponentially light and the effective theory will eventually break down. However, the point at which these effects become relevant for inflation will depend on the rate at which the states become light, which is parametrised by $\lambda$ in eq.~\eqref{bound}. From a geometric point of view, the concrete value of $\lambda$ can in principle depend on the type of trajectory followed in the field space. Any attempt to give a universal value  for $\lambda$  will imply to constrain both the geometry of the scalar manifold as well as the type of scalar potential that can arise from string theory, thus effectively constraining  the type of trajectories allowed by quantum gravity. This is a very difficult question since it would imply to know the full potential including all possible backreaction effects, to determine the final trajectory followed in the field space. However, we can aim at least to determine the value of $\lambda$ associated to each trajectory, even if we do not know what trajectories will be eventually allowed by quantum gravity. The advantage of using the Nilpotent Orbit Theorem to determine the \Kahler potential is that we can avoid part of the path dependence issues, as we will see below. We will distinguish a few cases depending on their level of technical difficulty.

\subsection{Saxionic trajectories}

\begin{figure}[htb]
%\vspace*{3mm}
%\hspace{-3mm}
\begin{center}
\includegraphics[width=8.5cm]{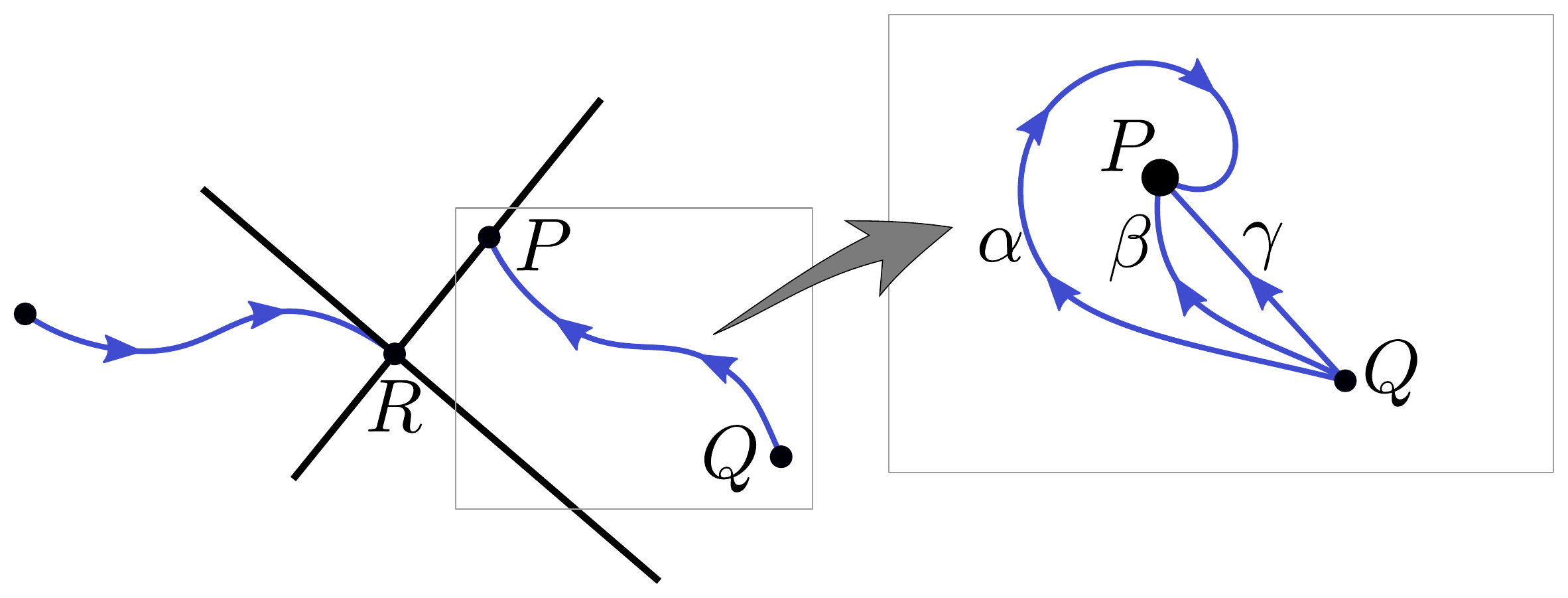}
\end{center}
%\vspace*{-0.3cm}
\caption{Sketch of different trajectories in the moduli space. The straight black lines correspond to singular divisors, while the blue lines correspond to different type of trajectories approaching the singular loci. The left figure refers to whether the trajectory is approaching a one-parameter (point $P$) or a multi-parameter degeneration (point $R$). Notice that, even if each singular divisor has one transverse complex coordinate $T_i$, the left drawing corresponds to the plane spanned only by the two real components $\phi_i$. The right figure corresponds instead to the complex plane spanned by $T_1=\phi_1+i\theta_1$. It can be therefore seen as a transversal view of the left figure so that one of the singular divisors appears as a point. Different trajectories $\alpha,\beta,\gamma$ correspond to different mixings of the saxion $\phi_1$ with the axionic field $\theta_1$, so that $\gamma$ is a purely saxionic trajectory.}
\label{FIG:curva}
\end{figure}

Let us still assume we are moving only along saxionic fields, i.e. on radial directions to the singularity. In a higher dimensional moduli space, with multiple saxions, we can distinguish two cases depending on the number of transverse coordinates to the singular point.
\begin{itemize}
\item One-parameter degenerations: the singular point is located at a single singular divisor at $\phi\rightarrow \infty$. In other words, there is only one transverse complex coordinate $T$ to the singularity and, therefore, only one radial way $\phi$ to approach it. 
\item Multi-parameter degenerations: the singular point is located at the intersection of multiple singular divisors, each of them located at $\phi_i\rightarrow \infty$ where $i$ runs over the number of singular divisors. In other words, there is more than one complex coordinate transverse $T_i$ to the singular point.
\end{itemize}
These two cases have been schematically shown in fig.~\ref{FIG:curva}, where the point $P$ is a one-parameter degeneration while $R$ is a multi-parameter degeneration as corresponds to the intersection of two singular divisors.

It is important to remark that, in this paper, we have only investigated the cosmological implications of one-parameter degenerations. The \Kahler potential was given in eq.~\eqref{Ks}. For multi-parameter degenerations, the polynomial in $e^{-K}$ will be an expansion over each transverse coordinate. Interestingly, there is a very rich underlying structure of the possible enhancements and intersections which will again constrain the physics and geometry around these points. This has been analysed in \cite{Grimm:2018cpv}. Even if path dependence issues become more important, it is possible to classify paths into different growth sectors and to determine
how the masses of the tower of states behave within each growth sector. A more detailed analysis of the field metrics in these growth sectors is left for future work.

Even if we restrict ourselves to one-parameter degenerations, there can also be additional scalar fields $z_a$ parametrizing the longitudinal directions to the singular divisor. All the bounds obtained for the single field case still apply in this case, with the difference that the constant $d$ appearing in the leading term of the kinetic term for the saxion is replaced by some $d_{\text{eff}}$ which will also depend on the other spectator fields $z_a$. However, $d_{\text{eff}}$ is upper bounded by the integer $d$, i.e.  $d_{\text{eff}}\leq d$, so the bounds on the field range for purely saxionic trajectories become even stronger. %The result for trajectories involving some axionic component will work similarly than in the previous subsection.

\subsection{Saxion-Axionic trajectories}

Let us finally consider the case in which we also displace some axionic field. 
 Varying the saxion is equivalent to moving along the radial direction towards the infinite distance singularity, while varying the axion corresponds to circle around the singularity. In the previous section, we only considered purely saxionic trajectories, as in the original models of $\alpha$-attractors. Here we will comment on the bounds for more general trajectories in which we also move along the axionic direction. Different trajectories have been depicted in fig.~\ref{FIG:curva} (right figure). For simplicity, we will assume that the moduli space is only parametrised by one complex scalar field $T=\phi+i\theta$. The leading term for the kinetic field metric near the infinite distance singularity reads
\beq
\mathcal{L}=-\frac{d}{4\phi^2}(\partial \phi)^2-\frac{d}{4\phi^2}(\partial \theta)^2\,.
\eeq
It can be proved \cite{wang1} that any real smooth trajectory approaching the point $\phi\rightarrow \infty$ has infinite length,
\beq
\Delta \varphi\gtrsim \frac{\sqrt{d}}{2}\log(\phi)\rightarrow \infty\,,
\eeq
even if it has an axionic component.
In particular, a trajectory involving a linear combination of saxion and axion such that 
\beq \label{trajectory}
(\phi,\theta)=(\phi_0+\delta\phi,\frac{1}{a}\delta\phi)
\eeq
has a length given by
\beq
\Delta\varphi =\frac{ \sqrt{d(1+a^2)}}{2a}\log(\phi_0+\delta\phi)\,.
\label{distanceaxion}
\eeq
However, it is not enough to know the field metric in order to determine the drop-off of the cut-off and the maximum field range consistent with the conjecture. We also need information about the mass behaviour  of the tower of particles. For a single complex field, it is still expected that the mass scales at leading order as
\beq
m\sim \frac{m_0}{\phi^p}\,.
\eeq
Plugging eq.~\eqref{distanceaxion} into the species bound for this tower of particles, and denoting $\lambda=2p/\sqrt{3d}$ as before, we get
\beq
\Delta\varphi\leq \Delta\varphi_{\text{SDC}} =\frac{1}{\lambda}\frac{\sqrt{1+a^2}}{a}\log\frac{M_p}{H}\,.
\eeq
This is the same formula used in the previous analysis but with a new effective factor $\lambda_{\text{eff}}=\lambda\frac{a}{\sqrt{1+a^2}}$ which now depends on how much we are moving along the axionic direction. For $a\rightarrow \infty$, the trajectory is mainly saxionic and we recover the results of section \ref{sec:swampland_const}. The field range cannot be made parametrically large and it is upper bounded by $\lambda^{-1}$. For $a\rightarrow 0$, the trajectory is mainly axionic and in principle $\lambda_{\text{eff}}\ll \lambda$ so $\Delta\varphi_{\text{SDC}}$ can be made parametrically larger. This limit is equivalent to consider an axion monodromy inflationary model \cite{Silverstein:2008sg,McAllister:2008hb} in which backreaction effects involve some displacement of the saxion fields which is quantified by the value of $a$. If $a$ is very small means that backreaction effects are negligible and the trajectory is mainly axionic. For this to happen, one needs to be able to engineer a mass hierarchy such that the axion is much lighter than the saxion \cite{Valenzuela:2016yny}. The question, though, is whether there exists a potential in an effective theory consistent with  quantum gravity that allows us to generate this mass hierarchy and move along this almost purely axionic trajectory. In many cases, it turns out that the same tuning required to get this mass hierarchy in string compactifications, also brings the effective theory away from the perturbative controllable regime \cite{Blumenhagen:2017cxt,Blumenhagen:2018nts} (see though \cite{Landete:2017amp}). Therefore, it remains as an open challenge to fully globally engineer a controllable model of this type \footnote{See, though, the recent work \cite{Buratti:2018xjt} where the backreacted kinetic term along the axionic trajectory goes as $1/\delta\phi$, seemingly  implying a parametrically large axionic field displacement. However, in this case, the axion varies over a non-compact spatial dimension.}. All we can say, though, is that this type of monodromic axionic trajectories are still the best candidates to generate larger field excursions as of now. %The mass hierarchy between axion and saxion required to get this tuned trajectory has not been able to been realised yet in a controlled string compactification.

Regarding the cosmological predictions, in the case of a mixed saxion-axion trajectory, asymptotic to eq.~\eqref{trajectory} at infinite distance, one would generically still expect an emergent inflationary behaviour typical of $\alpha$-attractors (provided the conditions on the potential discussed above). In fact, the  kinetic term has still an effective $1/\phi^2$ dependence along that path. In the ideal case of perfect linearity, with $a$ being a constant along the whole trajectory, the inflationary predictions will be again given by eq.~\eqref{trajectory} but with $\alpha$ being replaced by $\alpha_{\text{eff}}$ such as
\beq
\alpha\ \rightarrow\ \alpha_{\text{eff}}= \alpha\ \frac{1+a^2}{a^2}\,.
\eeq
For small $a$, that is when the axionic component becomes relevant, the tensor-to-scalar ratio will therefore increase. 

However, moving away from the singularity, deviations from the linear saxion-axion combination are generically expected. These will depend on the specific details of the full scalar potential\footnote{Deviations from the linear case eq.~\eqref{trajectory}  can be again encoded in the trajectory-parameter with a dependence such as $a=a_0+a_1/\phi + \mathcal{O}(1/\phi^2)$.}, with direct consequences on the resulting inflationary dynamics. The evolution along the axionic direction  can in fact lead to phenomenologically distinct scenarios. Here below, we give an account of quite generic situations, keeping in mind that many model-dependent subtleties might come into play.

One  possibility is that the parameter $a$ effectively increases thus yielding a purely saxionic trajectory.  If the potential is such that the last 60 e-foldings of inflation happen along this path, then we recover the original predictions given by eq.~\eqref{trajectory}.  On the contrary, if the trajectory becomes mainly axionic (effective reduction of $a$) and if the potential is suitable to sustain again at least  60 e-foldings of quasi-exponential expansion, then the predictions will be strictly dictated by the form of the axionic potential (with model dependent outcomes). In fact, CMB observations will `see' just the last stage of inflation, thus ignoring the preceding cosmological history.

Another possibility is that the first stage of inflationary attractor along the diagonal is interrupted with a subsequent phase, which can sustain an insufficient period of inflation or even no inflation at all (thus leading to a premature ending of inflation). In both cases, the cosmological predictions will read
\begin{equation}
n_s= 1-\frac{2}{\left(N+\delta N\right)}\,, \qquad r=\frac{12\ \alpha_{\text{eff}}}{\left(N+\delta N\right)^2}\,,
\end{equation}
with $N\approx 60$ and $\delta N$ being a positive quantity in case the inflationary period is prematurely arrested (see e.g. \cite{Roest:2016lrb}), while it is a negative number if there is a second stage of insufficient inflation, with a duration of $|\delta N|<60$ e-foldings (see e.g. \cite{Christodoulidis:2018qdw,Linde:2018hmx}).

\section{Conclusions}

In this paper we have discussed the interplay between the Swampland Distance Conjecture, the physics at infinite distance singularities and its implications for inflation. Specifically, we have pointed out that the emergent field metric predicted by the SDC (and confirmed by explicit analysis in the context of Calabi-Yau manifolds \cite{GPV}) is the same as the typical of $\alpha$-attractor models of inflation.  The SDC therefore suggests that any scalar field travelling along a {\it non-compact} trajectory towards a boundary of the string moduli space will exhibit the phenomenology of $\alpha$-attractors for large enough parametrically field values, provided a certain regularity on the scalar potential (see Sec.~\ref{Sec:alpha}).  However, the conjecture implies also that the limit where the  universal inflationary behaviour emerges is the same where the effective field theory is supposed to break down cause of the appearance of an infinite tower of massless states. We have investigated these aspects and found a number of interesting results:
% that, although the infinite long plateau typical of $\alpha$-attractors is forbidden by the SDC, these models can still deliver a sufficient amount of inflation ($N>60$) with cosmological predictions equal to eq.~\eqref{nsr}. For $\alpha$-attractor models arising at infinite distances, the upper bound on the total number of e-foldings is given by eq.~\eqref{Nbound}.

%In order to find our final results, a number of intermediate steps have been crucial:
\begin{itemize}
\item We have first pointed out that assuming validity of the EFT of inflation ({\it i.e.} $H\leq\Lambda_{QG}$) and eq.~\eqref{bound1} automatically leads to a universal upper bound on the inflaton field range in terms of the tensor-to-scalar ratio $r$ measured at horizon exit. This bound scales as $\Delta\varphi \lesssim - \log(r)$ and, when compared with the Lyth bound ($\Delta\varphi \gtrsim\sqrt{r}$), it defines an area of inflationary models consistent with the quantum gravity constraints imposed by the SDC (see fig.~\ref{FIG:bounds}).
 
\item We have argued that if $\alpha$-attractors are realised within string theory and by means of only one radial transverse direction evolving during inflation, then the kinetic metric eq.~\eqref{kin1} should vanish in the deep inflationary limit. This happens because the physics emerges around infinite-distance singularities in field space. In Calabi-Yau string compactifications, the symmetries of these spaces typically do not include the inversion transformation, which makes this inflationary construction intrinsically different from the {\it pole inflation} scenario \cite{Galante:2014ifa,Broy:2015qna,Dias:2018pgj} (although with identical predictions). Interestingly, this realization comes together with the additional feature that the decay constant of the axion partner goes to zero in this limit, thus implying consistency also with the bounds imposed by the WGC. The increase of the axion decay constant towards the end of inflation might lead to interesting phenomenological consequences. 

\item We have proven that, although the infinite long plateau typical of $\alpha$-attractors is forbidden by the SDC, these models can still deliver a sufficient amount of inflation ($N>60$) with cosmological predictions equal to eq.~\eqref{nsr}. For $\alpha$-attractor models arising at infinite distances, the upper bound on the total number of e-foldings is given by eq.~\eqref{Nbound}.

\item The realization of \mbox{$\alpha$-attractors} at infinite distances, allowed us to relate, through eq.~\eqref{dalpha}, the parameter $\alpha$ of these inflationary class of models with the parameter $d$, the latter being intimately related to the fundamental geometric properties of the singularity (see eq.~\eqref{Ks} and eq.~\eqref{kinSDC}). For saxionic trajectories,  $d$ is upper bounded by an integer depending on the properties of the monodromy transformation of infinite order around this point (which translates to the shift symmetry of the axion partner in the effective theory), thus implying an upper bound eq.~\eqref{boundr} on the tensor-to-scalar ratio. It is also eventually related to the rate at which the infinite tower of particles becomes light, as both parameters are connected to the asymptotic structure of the kinetic term. More concretely, we get $\alpha\sim \lambda^{-2}$ where $\lambda$ appears in the definition of the SDC eq.~\eqref{bound1}. A bound on $\lambda$ has, therefore, a direct impact on $\alpha$-attractors.

\end{itemize}

Many of the results discussed in this paper are (model-independent) consequences of applying the Nilpotent Orbit Theorem  \cite{schmid} to the \Kahler potential of Calabi-Yau compactifications. Therefore, the validity of these conditions is restricted to these spaces, although their deep relation to the SDC suggests that they might be valid in general. It is important to remark, though, that they have only been proven for one-parameter degenerations, meaning that there is only one saxion going to infinity while all the other fields are kept finite. Even if we expect them to be true in general, we cannot apply yet our results to models like {\it fibre inflation} \cite{Cicoli:2008gp},  which involves  two fields moving towards a singular point located at the intersection of multiple singular divisors (see e.g. point $R$ of fig.~\ref{FIG:curva}). Although we have already outlined some of the generic expectations of multi-parameter degenerations in Sec.~\ref{multifield} , we leave this interesting but more involved analysis  for future work.

Finally, we want to remark that much more effort is still required to prove the conjecture and determine the parameter $\lambda$ in eq.~\eqref{bound1} from first principles. This is essential in order to ever give precise constraints on inflation which aim to be universal.  
Interestingly, the mathematical structure underlying the infinite distance singularities can also be used to potentially constrain the asymptotic structure of the scalar potential in flux string compactifications. This would represent an essential complementary analysis, since in the present work we have focused just on the kinetic structure of the inflaton field. We hope to return soon to these interesting topics.

 %\vspace{1.5cm}

\acknowledgments
We would like to thank Thomas Grimm, Liam McAllister, Miguel Montero, Fernando Quevedo, Matthew Reece, Yvette Welling and Timm Wrase  for helpful comments and discussions. MS is supported by the Research Foundation - Flanders (FWO) and the European Union's Horizon 2020 research and innovation programme under the Marie Sk{\l}odowska-Curie grant agreement No. 665501. IV is supported by the Simons Foundation Origins of the Universe program (Modern Inflationary Cosmology collaboration).

%\newpage

%\begin{thebibliography}{99}

%%%%%%%%%%%%%%%%%%%%%%%%%%%%%%%%
%%%%%%%%%%%%Bibliography%%%%%%%%%%%%%%
%%%%%%%%%%%%%%%%%%%%%%%%%%%%%%%%

\bibliography{RefSDC}
\bibliographystyle{utphys}

\end{document}